# High-excitation molecular gas in local luminous AGN hosts

Padeli P. Papadopoulos[1,2], A. Kovacs[3], A. S. Evans[4,5], and P. Barthel[6]

[1] Argelander Institut für Astronomie, Auf dem Hügel 71, 53121 Bonn, Germany
e-mail: padeli@astro.uni-bonn.de
[2] Institute of Astronomy & Astrophysics, National Observatory of Athens, I. Metaxa & V. Paulou, P. Penteli, 15236, Athens, Greece
[3] Max-Planck-Institut für Radio Astronomie, Auf dem Hügel 69, 53121 Bonn, Germany
e-mail: akovacs@mpifr-bonn.mpg.de
[4] Department of Physics and Astronomy, State University of New York at Stony Brook, NY 11794-3800, USA
e-mail: aaron.evans@stonybrook.edu
[5] Department of Astronomy, University of Virginia, P. O. Box 400325, Charlottesville, VA 22904 and National Radio Astronomy Observatory, 520 Edgemont Road, Charlottesville, VA 22903
[6] Kapteyn Institute, University of Groningen, P.O. Box 800, 9700 AV Groningen, The Netherlands
e-mail: pdb@astro.rug.nl

**ABSTRACT**

*Aims.* The discovery of large amounts of molecular gas in the hosts of luminous AGN in the local Universe offers a unique opportunity to study its physical conditions under the influence of starbursts and powerful AGN in great detail with present and future mm/sub-mm telescopes. Towards achieving this goal we present first results of multi-J CO line observations of such objects.
*Methods.* We made use of the mm/sub-mm receivers on the James Clerk Maxwell Telescope (JCMT) to conduct observations of the CO J=3–2, 2–1 lines in five local, optically powerful AGN and of the higher excitation CO J=4–3 line in 3C 293 (a powerful radio galaxy).
*Results.* Luminous CO J=3–2 emission and high CO (3–2)/(1–0) intensity ratios are found in all objects, indicating highly excited molecular gas. In 3C 293, an exceptionally bright CO J=4–3 line with a high CO(4–3)/(1–0) ratio cannot be easily explained given its quiescent star-forming environment and very low AGN X-ray luminosity.
*Conclusions.* In 3C 293, shocks emanating from a well-known interaction of a powerful jet with a dense ISM may be responsible for the high excitation of its molecular gas on galaxy-wide scales. Star formation can readily account for the gas excitation in the rest of the objects, although high X-ray AGN luminosities can also contribute significantly in two cases. Measuring and eventually imaging CO (high-J)/(low-J) line ratios in local luminous QSO hosts can be done by a partially completed ALMA during its early phases of commissioning, promising a sensitive probe of starburst versus AGN activity in obscured environments at high linear resolutions.

## 1. Introduction

The discovery of large molecular gas masses in optical and radio-selected quasars in the local Universe (Alloin et al. 1992; Evans et al. 1999a, 2001, 2005; Scoville et al. 2003; Bertram et al. 2007) has established the presence of abundant fuel for the intense star-formation activity in their hosts (Haas et al. 2000; Barthel 2001, 2006). However, not all that molecular gas belongs to the dense and warm phase which directly fuels star formation (e.g. Gao & Solomon 2004), while AGN can raise the *global* molecular line excitation via their often high X-ray luminosities (Meijerink & Spaans 2005; Meijerink 2006), complicating the use of the molecular gas excitation state as a unique marker of star-forming activity in AGN+starburst composite systems.

Using the HCN J=1–0 line luminosity to estimate the dense star-forming molecular gas mass in such systems is hindered by its faintness (I(HCN)/I(CO) ≲ 1/10 for J=1–0), making its detection for a significant number of local AGN hosts with current instrumentation challenging, even for gas-rich systems (e.g. Evans et al. 2005, 2006). It becomes a true struggle for high-J HCN line observations, necessary if the HCN(1-0)-luminous gas is to be safely attributed to a dense *and* warm star-forming phase. On the other hand high-J CO transitions are much brighter and easier to detect in Luminous Infrared Galaxies (LIGs, $L_{8-1000\,\mu m} > 10^{11}\,L_\odot$) in a systematic fashion and, barring any AGN-related boosting, CO (high-J)/(low-J) intensity ratios have been shown to be good tracers of star-forming molecular gas within galaxies (Yao et al. 2003; Zhu et al. 2003). Finally, the large number of high-J CO transitions detected at high redshifts (e.g. Weiss et al. 2007) makes the CO spectral line energy distributions (SLEDs) the first molecular line benchmarks of choice for comparisons between local LIGs/AGN and their high redshift counterparts.

We report the first results of multi-J CO line observations of luminous AGN hosts in local Universe. Throughout this work we assume a cosmology with $H_\circ = 71\,\text{km s}^{-1}\,\text{Mpc}^{-1}$, $\Omega_M = 0.27$, $\Omega_\Lambda = 0.73$.

## 2. The sample and the observations

The observations were conducted during several sessions from 1999 up to 2005, as a pilot extension of a large molecular line survey of local LIGs (see Papadopoulos et al. 2007 for details) towards the hosts of five optically luminous local AGN that also had CO J=1–0 detections reported in the literature (to en-able molecular gas excitation analysis). All objects are optically-selected except 3C 293, a powerful radio galaxy included in the sample as a local representative of this AGN class (and the only such AGN with a CO J=1–0 detection at the time of our observations).

We used the B 3 (315-370 GHz, tuned single sideband), and the A 3 (211-276 GHz, double sideband) receivers at the 15-meter James Clerk Maxwell Telescope (JCMT) on Mauna Kea (Hawaii) to conduct sensitive observations of the CO J=3–2, 2–1 transitions ($\nu_{32}$ = 345.796 GHz, $\nu_{21}$ = 230.530 GHz). Typical system temperatures were $T_{sys}$(A 3) $\sim$ (350–390) K and $T_{sys}$(B 3) = (370 − 490) K (including atmospheric absorption), except for PG 0050+124 (I Zw 1) where CO J=3–2 is redshifted to $\sim$ 325.9 GHz (near an atmospheric absorption band) where $T_{sys}$(B 3) $\sim$ 800 K. The CO J=4–3 line ($\nu_{43}$ = 461.041 GHz) towards 3C 293 was observed in 18 November 2004 and 23 April 2005 using the now decommissioned receiver W operating in C-band (430-510 GHz), under very dry conditions that yielded $T_{sys}$(W/C) $\sim$ (1600 − 2400) K.

In all cases the Digital Autocorrelation Spectrometer (DAS) with 920 MHz or 1.8 GHz bandwidth was the backend, and rapid beam switching at 1 Hz and a beam throw of 60″ − 120″ (in azimuth) ensured exceptionally flat baselines. In the case of PG 1700+518 the CO J=3–2 line has been independently detected in two observing sessions three years apart. We co-added the edited spectra and subtracted linear baselines (see Figure 1). The velocity-integrated line flux densities were then estimated from those spectra using

$$S_{co} = \int_{\Delta V} S_\nu \, dV = \frac{8k_B}{\eta_a \pi D^2} K_c(x) \int_{\Delta V} T_A^* \, dV =$$
$$= \frac{15.6 (Jy/K)}{\eta_a} K_c(x) \int_{\Delta V} T_A^* \, dV, \quad (1)$$

where $K_c(x) = x^2/(1 − e^{-x^2})$, $x=\theta_s/(1.2\theta_{HPBW})$ ($\theta_s$=source diameter) accounts for the geometric coupling of the gaussian part of the beam with a finite-sized, disk-like source, and $\eta_a$ is the corresponding aperture efficiency. For A 3 and B 3 receivers the latter was checked using planet observations, and found consistent with the average values reported in the JCMT efficiencies database[1], during the specific periods of our observations, within the nominal $\sim$ 15% calibration uncertainty.

For $\eta_a$(W/C) no large database is available, and thermal/elevation-dependent dish deformation can result to large variations of its value, even on prime sub-mm telescopes like the JCMT. We thus adopt the average of our own measurements $\eta_a$(W/C) = 0.32 ± 0.06 (the error is the dispersion of our measurements), obtained from observations of Mars and Uranus during the same days as those of 3C 293 and at similar elevations. The beam sizes were $\theta_{HPBW}$(230-268 GHz)=18″-21″, $\theta_{HPBW}$(B 3)=14″ and $\theta_{HPBW}$(W/C)=11″. The telescope pointing (crucial for high-frequency observations of compact sources), checked hourly using strong spectral line or continuum sources, was found accurate to $\sim$ 1.5″-3″(rms) at 230, 345 GHz, and $\sim$ 3.5″ (rms) at 461 GHz.

The relevant observational data are in Table 1, while the velocity-integrated line fluxes, CO J=1–0 data from the literature and the corresponding line ratios, are in Table 2. The measurement errors consist of the thermal rms error and a $\sim$ 15% calibration uncertainty (estimated from strong spectral line standards observed during our runs). Only for Mrk 1048 the CO source size is larger than the JCMT beam and thus even the corrected CO J=3–2 line flux (and thus the reported CO (3–2)/(1–0) ratio) is a lower limit. For PG 1119+120 while no single-dish CO J=1–0 line measurement exists, a compact H-band size of $\sim$ 1.39 kpc ($\sim$ 1.44″ at z=0.050) (Veilleux et al. 2006) ensures that its total molecular gas reservoir is accurately depicted by the interferometric CO 1–0 image. For PG 1700+518 where no such CO image exists, imaging at H, K bands reveals very compact emission ($\lesssim$ 2.5″, Márgez et al. 2001) and thus both CO J=1–0 and J=3–2 measurements are expected to reflect total line flux.

## 3. Highly excited molecular gas in AGN hosts: starburst versus AGN

CO J=3–2 is detected in all objects (and J=2–1 in the three that were observed in this transition), and in all but one (3C 293) we find CO (3–2)/(1–0) line ratios of $r_{32} \sim 0.5 − 1.0$ (Table 2), typical for star-forming galactic nuclei and LIGs (e.g. Devereux et al. 1994; Yao et al. 2003; Narayanan et al. 2005). A critical density of $n_{crit}(3 − 2) \sim 10^4$ cm$^{-3}$, and $E_3/k_B$ = 33.2 K makes high $r_{32}$ values excellent tracers of the warm and dense gas found in star-forming regions. Noting that even in M 82, a nearby starburst, a massive quiescent gas phase with expected $r_{32} \lesssim 0.3$ dominates over that fueling its star formation (Weiss, Walter, & Scoville 2005), sets the hereby observed high *global* $r_{32}$ ratios in AGN hosts in perspective. Nevertheless, the presence of powerful AGN makes a simple starburst interpretation of the observed high gas excitation far from straightforward.

### 3.1. A luminous CO J=4–3 line in 3C 293: shock-excited molecular gas?

This powerful radio galaxy was the first local F-R II source (Fanaroff & Riley 1973) detected in CO (Evans et al. 1999b, 2005), and its large linewidth (FWZI$\sim$ 850 km s$^{-1}$), disturbed morphology (Floyd et al. 2006), and substantial molecular gas reservoir of $\sim 5 \times 10^9$ M$_\odot$ (estimated from its CO J=1–0 line luminosity and the CO-H$_2$ conversion factor from Downes & Solomon 1998) puts it on par with LIGs. Yet quite unlike them it has a star formation efficiency of L$_{IR}$/M(H$_2$) $\sim$ 8 L$_\odot$/M$_\odot$ typical of the quiescent ISM in spirals (Evans et al. 1999b, 2005), and a rather low star formation rate of $\sim$ (6 − 7) M$_\odot$ yr$^{-1}$ (estimated from its far-IR luminosity). Thus, its luminous CO J=4–3 line with $\int S_{CO}(4-3)\, dV$=(842 ± 253) Jy km s$^{-1}$ and high CO (4–3)/(1–0) ratio of $r_{43}$ = 1.0 ± 0.34 ($> r_{32}, r_{21}$) comes as a surprise, and clearly indicates *the presence of an additional high-excitation gas phase, unhinted by the three lower-J CO transitions or the level of its star formation activity*.

The excitation characteristics of the J=4–3 line ($E_4/k_B \sim$ 55.3 K, $n_{crit} \sim 1.9 \times 10^4$ cm$^{-3}$), make even the lowest possible value of $r_{43} \sim$ 0.54 (for $K_{c,43}(x)$ = 1, and $(T_{43} − \sigma)/(T_{10} + \sigma)$) typical of warm and dense molecular gas (e.g. Güsten et al. 1993; White et al. 1994), while the quiescent ISM environment of the Galaxy by comparison has $r_{43} \sim$ 0.10−0.17 (Fixsen et al. 1999). High $r_{43}$ ratios have been measured in powerful radio galaxies before but for high redshift systems where tremendous starbursts make a large mass of highly excited molecular gas an imperative (e.g. Greve et al. 2004; Papadopoulos et al. 2005). Given that $r_{43} > r_{32}, r_{21}$, our Large Velocity Gradient (LVG) radiative transfer code, used to constrain the gas physical conditions in 3C 293, predictably fails to account for its luminous CO J=4–3 emission with a single gas phase. Single-phase fits can be barely found only for $r_{32,21} + \sigma$ *and* $r_{43}$(min) = 0.54 where $T_k \sim 40-85$ K and $n(H_2) \sim 10^3$ cm$^{-3}$, while fitting only the $r_{43}$ ratio yields $n(H_2) \sim (10^4 - 3 \times 10^5)$ cm$^{-3}$ for $T_k \geq T_{dust} \sim 40$ K.

---
[1] http://www.jach.hawaii.edu/JCMT/spectral_line/Standards/eff_web.html

Large X-ray luminosities, frequently found in powerful radio-loud AGN such as 3C 293, can in principle boost mid/high-J CO line fluxes by keeping the gas warm (and thermally de-coupled from a presumably cooler dust reservoir) much deeper into molecular clouds than FUV photons (Meijerink, Spaans, & Israel 2006). However 3C 293 stands out as one of the few powerful radio galaxies in the local Universe with a very low X-ray luminosity of $L_{0.1-2.4\,keV} \leq 1.15 \times 10^{41}$ erg s$^{-1}$ ($\sim 3 \times 10^7$ L$_\odot$, Miller et al. 1999). At distances of $\sim (0.86-3)$ kpc which encompass the bulk of its molecular gas reservoir (Evans et al. 1999b) the expected X-ray flux of $F_x \leq (9 \times 10^{-5} - 1.3 \times 10^{-3})$ erg cm$^{-2}$ s$^{-1}$ corresponds to an equivalent $G_{X,\circ} \lesssim 0.06-0.8$ of effective energy flux (in Habing units). This is comparable, at best, to the Galactic interstellar radiation field and thus incapable of creating large scale X-ray Dissociation Regions (XDRs). Hard X-rays have been recently detected in 3C 293 with $L_{2-10\,keV} = 7 \times 10^{42}$ erg s$^{-1}$ (Hardcastle 2008). This yields $G_{o,X} = 3.65-48.7$, making a "soft" XDR possible, though the major uncertainty affecting the value of $G_{X,o}$ (the size of the molecular gas reservoir) remains.

Turbulent heating from the dissipation of shocks can produce highly excited gas while the dust temperature (and thus the IR luminosity) remain modest. This happens for the dense and highly turbulent molecular clouds in the Galactic center where gas with $T_k \sim 150$ K (Rodriguez-Fernández et al. 2001) is concomitant with cool dust of $T_{dust} \sim 20$ K (Pierce-Price et al. 2000). In 3C 293 a very powerful jet (Floyd et al 2006) drives an impressive large scale HI outflow via its interaction with the molecular gas (Morganti et al. 2003), and large scale shocks from such an interaction may cause the high molecular gas excitation observed in this system. This would make it the first known case of galaxy-wide shock-excited molecular gas, and higher-J CO line and hard X-ray (E > 2.5 keV) observations will be crucial in confirming this.

### 3.2. Physical conditions of the molecular gas: far-UV versus X-ray excitation

Except PG 1119+120, evidence for intense star formation along with large molecular gas mass reservoirs ($\sim (4-16) \times 10^9$ M$_\odot$) exists in the rest of the AGN hosts and can easily account for the high excitation state of their molecular gas. The well-studied QSO host PG 0050+124 (I Zw 1) is such a case, with clear evidence of an ongoing circumnuclear starburst (Schinnerer, Eckart, & Tacconi 1998). An LVG fit of all its CO line ratios (Table 1) along with a known $^{12}$CO/$^{13}$CO J=1–0 ratio of $R_{10} = 9 \pm 2$ (Eckart et al. 1994) yields $T_k \sim 65$ K and $n(H_2) \sim 10^5$ cm$^{-3}$, contained within a wider range of LVG solutions with $T_k \sim (35-75)$ K and $n(H_2) \sim (10^4 - 3 \times 10^5)$ cm$^{-3}$. Similar conditions, typical of star forming gas, correspond to the line ratios of the rest of the AGN hosts while their star formation efficiencies of $L_{FIR}/M(H_2) \sim (50-100)$ L$_\odot$/M$_\odot$ are similar to those of the vigorously star-forming LIGs. However AGN with high X-ray luminosities exist in PG 0050+124 ($L_{0.5-2\,keV} = 3.5 \times 10^{10}$ L$_\odot$; Piconcelli et al. 2005) and Mrk 1048 ($L_{2-10\,keV} = 10^{12}$ L$_\odot$; Risaliti et al. 2000). For PG 0050+124, at the distance of $\sim 1.75$ kpc where its star-forming molecular gas ring is located, this corresponds to $\sim 0.6$ erg cm$^{-2}$ s$^{-1}$ or $G_{X,\circ} \sim 375$. For the much more X-ray luminous AGN in Mrk 1048 even at 10 kpc away a flux of $\sim 0.35$ erg cm$^{-2}$ s$^{-1}$, or $G_{X,\circ} \sim 220$, is expected. Thus in both these cases X-rays can be competitive with starburst-originating FUV photons in exciting their molecular gas reservoirs.

## 4. Conclusions

We report first results of a pilot multi-J CO molecular line observations of the hosts of luminous local AGN. These can be summarized as follows
1. Highly-excited molecular gas is found in all the AGN host galaxies observed.
2. An exceptionally bright CO J=4–3 line in the powerful radio galaxy 3C 293 reveals highly excited gas, inconspicious in the lower three CO transitions, and unexpected given the low star formation rate and AGN X-ray luminosity in this system. Shock-excited molecular gas resulting from a well known strong jet-(dense ISM) interaction in 3C 293 maybe responsible.
3. In two AGN hosts, PG 0050+125 (I Zw 1) and Mrk 1048, X-ray powerful AGNs can contribute substantially to the global molecular gas excitation along with their resident starbursts.

These results demonstrate the promise of measuring and eventually imaging CO lines at $\nu \gtrsim 345$ GHz in the hosts of luminous AGN in the local Universe as probes into the excitation mechanisms of molecular gas. At the hereby reported line luminosities such imaging observations can be done by a partially completed ALMA during the early phases of its operation.

## 5. Ackowledments

We thank the referee Jeremy Lim for pointing out X-ray Dissociation Regions as a possibility for large scale molecular gas excitation in galaxies hosting AGN.

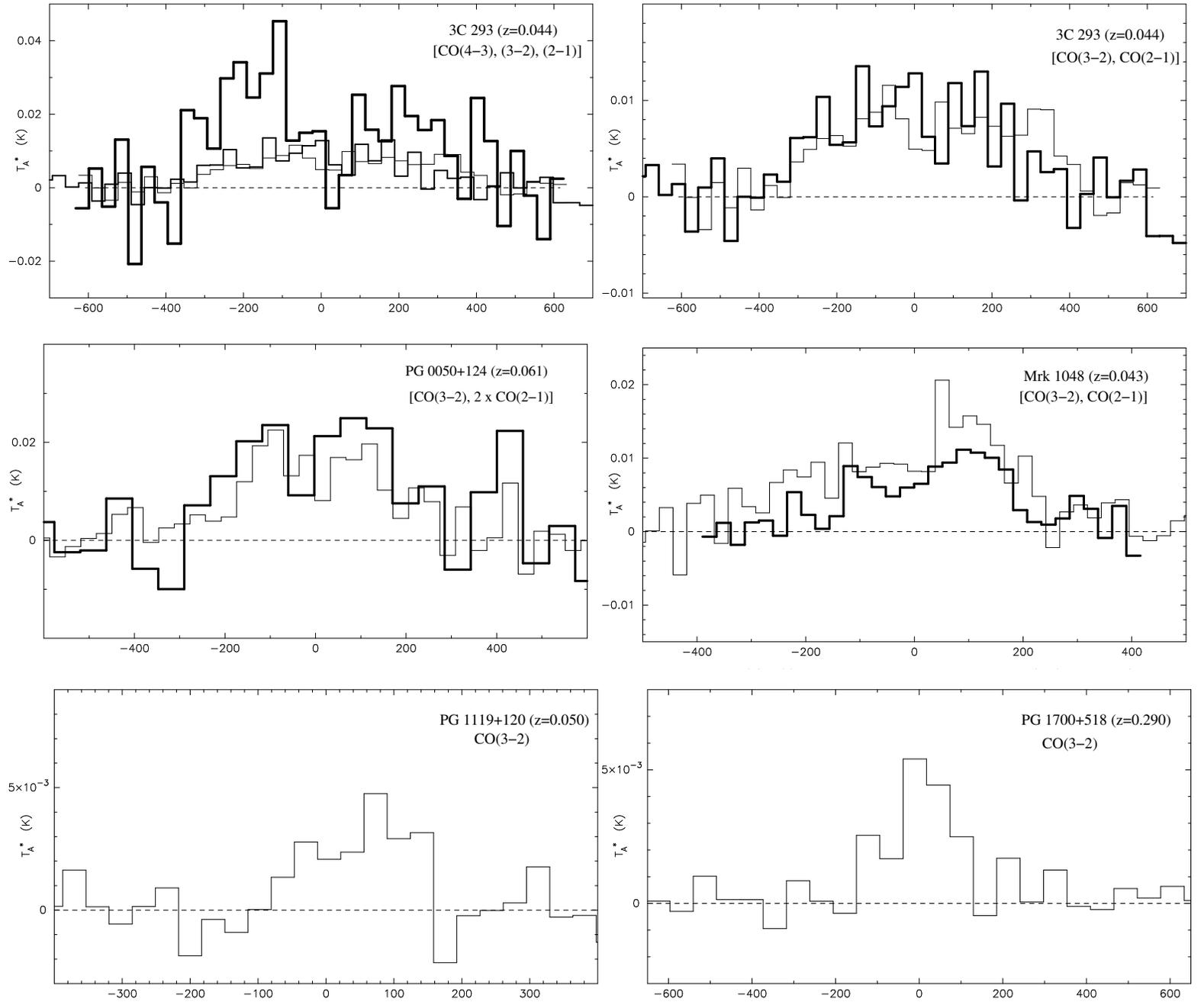

**Fig. 1.** The CO J=4–3 (for 3C 293 only), J=3–2 and J=2–1 spectral lines of the AGN hosts (the thicker the line the higher the rotational level), and the J=3–2 line in PG 1700+518, PG 1119+120. All spectra are in the $T_A^*$ scale (in PG 0050+124 the $^{12}$CO J=2–1 line has been scaled by a factor of 2 for clarity). The velocity resolution ranges from $\sim 25\,\mathrm{km\,s^{-1}}$ to $\sim 60\,\mathrm{km\,s^{-1}}$ and the $S/\sigma_{\mathrm{therm}}$ achieved for the velocity-integrated line intensity ranges from $\sim$5-6 (e.g. PG 1700+518, PG 1119+120) to $\geq 10$ (Mrk 1048).

**Table 1.** Observational parameters, and literature data

| Name (Type) | RA (B1950) | Dec (B1950) | $D_L(z)$ [a] | CO size [b] (Refs) | $L_{FIR}(L_\odot)$ (Refs) |
|---|---|---|---|---|---|
| PG 0050+124 (QSO) | $00^h\ 50^m\ 57^s.80$ | $+12° 25' 19.3''$ | 270 (0.061) | $\sim 12''$ (1,2) | $2.4 \times 10^{11}$ (7) |
| Mrk 1048 (Sy1) | $02^h\ 32^m\ 10^s.48$ | $-09° 00' 20.8''$ | 188 (0.043) | $\sim 22''$ (3,4) | $2.2 \times 10^{11}$ (4) |
| PG 1119+120 (QSO) | $11^h\ 19^m\ 10^s.93$ | $+12° 00' 45.4''$ | 219 (0.050) | $\leq 5''$ (5) | $5.0 \times 10^{10}$ (8) |
| 3C 293 (radio galaxy) | $13^h\ 50^m\ 03^s.15$ | $+31° 41' 32.5''$ | 192 (0.044) | $\sim 7''$ (6) | $3.75 \times 10^{10}$ (6) |
| PG 1700+518 (QSO) | $17^h\ 00^m\ 13^s.26$ | $+51° 53' 36.1''$ | 1482 (0.290) | ... | $9.6 \times 10^{11}$ (7) |

[a] Luminosity distance (Mpc), and redshift used for receiver tunnings.
[b] Size of the CO(1–0)-emitting region from interferometric images in the literature. (1,2)=Eckart et al. 1994, Schinnerer et al. 1990. (3,4)=Horellou et el. 1995, Appleton et el. 2002 (5,6)=Evans et al. 2001, 1999b; (7)=Haas et al. 2003; (8)=Bonatto & Pastoriza 1997

**Table 2.** CO line intensities and ratios

| Name | CO J=3–2 ($\eta_a$, $K_c(x)$)[a] (Jy km s$^{-1}$) | CO J=2–1 ($\eta_a$, $K_c(x)$)[a] (Jy km s$^{-1}$) | CO J=1–0 [b] (Refs) (Jy km s$^{-1}$) | $r_{21}$(CO) ($\frac{2-1}{1-0}$) | $r_{32}$(CO) ($\frac{3-2}{1-0}$) |
|---|---|---|---|---|---|
| PG 0050+124 | $356 \pm 107$ (0.55, 1.27) | $114 \pm 23$ (0.55, 1.11) | $34 \pm 7$ (1) | $0.84 \pm 0.24$ | $1.16 \pm 0.41$ |
| Mrk 1048 | $245 \pm 55$ (0.45, 2.087)[c] | $215 \pm 35$ (0.55, 1.387) | $49 \pm 6$ (2,3) | $1.10 \pm 0.22$ | $0.56 \pm 0.14$ |
| PG 1119+120 | $21 \pm 5$ (0.53, 1.0) | ... | $4.5 \pm 0.8$ (4) | ... | $0.52 \pm 0.15$ |
| 3C 293 | $208 \pm 55$ (0.45, 1.09) | $155 \pm 28$ (0.55, 1.035) | $52 \pm 8$ (5) | $0.74 \pm 0.17$ | $0.44 \pm 0.13$ |
| PG 1700+518 | $30 \pm 8$ (0.55, 1.0) | ... | $3.9 \pm 0.7$ (4) | ... | $0.85 \pm 0.26$ |

[a] The aperture efficiency, and the $K_c(x)$ factor used. The latter is estimated using CO(1-0) source sizes in Table 1. We set $K_c(x) = 1$ when no size information is available.
[b] CO J=1–0 from the literature (averages were used when more than one values were available. (1)=Barvainis et al. 1989 (for 4.5 Jy/K assumed for the IRAM 30-m at 115 GHz), (2,3)= Alloin et al. 1992, Appleton et al. 2002, (4)= Evans et al. 2008, (5)=Evans et al. 1999b